\documentclass[aps,prb,twocolumn]{revtex4}
\usepackage{epsfig}

\begin{document}
\headsep 2.5cm
\title{Validity of Anderson's theorem for s-wave superconductors.}

\author{Rostam Moradian$^{1,2}$} 
\email{rmoradian@razi.ac.ir} 
 \author{Hamzeh Mousavi$^{1}$}

\affiliation{$^{1}$Physics Department, Faculty of Science, Razi
University, Kermanshah, Iran\\
$^{2}$Institute for studies in Theoretical Physics and Mathematics (IPM)
            ,P.O.Box 19395-1795, Tehran, Iran}

\date{\today}

\begin{abstract}
  We investigate validity of Anderson's theorem (AT) for disordered s-wave superconductors in a negative U Hubbard model with random on-site energies, $\varepsilon_{i}$. The superconducting critical temperature, $T_{c}$, is calculated 
in the coherent potential approximation (CPA) as a function of impurity concentration, $c$,  the random  
potentials for different band filling.  In contradiction to Anderson's theorem, we found that $T_{c}$ is dramatically sensitive with respect to $c$ and $\varepsilon_{i}$. Our results shows that for low impurity concentrations and weak on-site energies, $\varepsilon_{i}$, the AT is valid, while in the strong scattering limit even for low impurity concentration, $T_{c}$ is very small with respect to the clean system and by increasing $c$ it is completely suppressed, hence AT is violated in this regime.

\end{abstract}

\maketitle
\section{ Introduction }

The insensitivity of the metallic conventional superconductors to nonmagnetic impurity doping is well known and was be explained  in terms of Abrikosov and Gorkov (AG) theory \cite{Abrikosov:59} and Anderson's theorem (AT)\cite{Anderson:59}. According to this theorem the superconductor critical temperature, $T_{c}$, and the superconductor density of states are not affected by the nonmagnetic impurity scattering. In contrast to this behavior in conventional metallic superconductors, in the high temperature superconductors (HTC) doping of a small amount of nonmagnetic impurities like $Zn$ destroys superconductivity, as reported for $YBa_{2}(Cu_{1-x}Zn_{x})_{3}O_{7}$\cite{Tokunaga:97}, $YBa_{2}(Cu_{1-x}Zn_{x})_{4}O_{8}$\cite{Itoh:01, Itoh2:01}, $La_{1.85}Sr_{0.15}Cu_{1-x}Zn_{x}O_{4}$\cite{Xiao:90}, $Bi_{2}Sr_{2}Ca(Cu_{1-x}Zn_{x})_{2}O_{8}$\cite{Kluge:95}. Physical evidence such as $T^{2}$ dependence of penetration depth in disordered HTC \cite{Hirschfeld:93, Bonn:94} and STM measurements on single $Zn$ doped on $Bi_{2}Sr_{2}Cu_{2}O_{8+\delta}$\cite{Pan:00}, shows that non-magnetic impurities behave as strong scattering centers.  While this can be understood as a consequence of d-wave pairing in HTC,
this kind of strong-scattering impurity role has raises the question for strong scattering effects in 
s-wave superconductors?  In particular, do the order parameter, density of states and especially $T_{c}$ change with impurity doping. In other words, is the Anderson theorem valid? Ghosal et al. \cite{Ghosal:98} calculated the local order parameter and the density of states of a disordered s-wave superconductor with a uniform impurity distribution, $[-V,V]$. They found that in the strong scattering limit, $V\rightarrow\infty$, for some regions of space the local order parameters are zero while in some others they are finite. A binary alloy s-wave superconductor has been investigated by one of us\cite{Moradian:01}, and it has been found that in the strong scattering limit the average order parameter goes to zero and a superconductor to insulator phase transition is take place, so the AT is violated in this case. Also the existence of an impurity bound state in a disordered s-wave superconductor energy gap is discussed by Balatsky et al.\cite{Balatsky:04}. 

In this paper we investigate the validity of AT for s-wave superconductors in detail, by calculating
 $T_{c}$ as a function of impurity concentration and electron band filling.
This is in contrast to previous work challenging AT in s-wave superconductors\cite{Moradian:01, Balatsky:04} which were calculations of the order parameter and density of states, not directly of $T_{c}$. 

 We found that at low impurity concentrations and in the weak scattering ($\varepsilon_{i}\;\langle\langle\; band\;width$) regime, the deviation of $T_{c}$ with respect to the clean system critical temperature  $T_{c0}$ is negligible, and hence AT is satisfied. But in low impurity concentrations at the strong scattering limit, although superconductivity is preserved, the $T_{c}$ is much less than $T_{c0}$, so the AT is not obeyed. By increasing $c$ in the strong scattering limit $T_{c}$ is completely suppressed hence a superconductor-insulator quantum phase transition take placed\cite{Moradian:01}. Furthermore, we also calculated $T_{c}$ 
 at a fixed averaged band filling for different impurity on-site energies and impurity concentration. Our results show that $T_{c}$  is decreased by increasing impurity concentration and eventually goes to zero in the strong scattering regime.

\section{Model and formalism.}
We start our investigations with the following attractive Hubbard model on a two dimensional square lattice\cite{Moradian:01},

\begin{equation}
H=-\sum_{ij}t_{ij}c^{\dagger}_{i\sigma}c_{j\sigma}+\sum_{i\sigma}
U_{i}\hat{n}_{i\sigma}\hat{n}_{i-\sigma}+ \sum_{i\sigma} (\varepsilon_{i}-\mu)
 \hat{n}_{i\sigma} .
\label{eq:Hamiltonian}
\end{equation}
where $t_{ij}$ are the hopping integrals, $U_{i}$ is the local
attractive interaction between electrons of opposite spin, and
$\mu$ is the chemical potential. Finally, $\varepsilon_{i}$ is the random
 on-site energy which takes values of $0$ with probability $c$ for impurity sites and $\delta$ with probability $1-c$ for the host sites.

In the mean field approximation, for any given configuration of
impurity sites the solution of Eq.~\ref{eq:Hamiltonian} is given
by the Bogoliubov de Gennes equation
\begin{equation}
\sum_{l} \left(
       \begin{array}{cc}
E\delta_{il} - H_{il}
 & \Delta_{i}\delta_{il} \\
\Delta^{*}_{i}\delta_{il} & E\delta_{il} +  H_{il}
\end{array}\right){\bf G}(l,j;E)=\delta_{ij}{\bf I},
\label{eq:Bogoliobovdegenneequation}
\end{equation}
where  $H_{il}= (\varepsilon_{i}-\mu_{i})\delta_{il}-t_{il}$ and
$\mu_{i}=\mu-\frac{1}{2}U_{i}n_{i}$ is the on-site
renormalized chemical potential due to the usual Hartree decoupling. The
local order parameter $\Delta_{i}$, assumed to be real, is defined by,
\begin{equation}
\Delta_{i}=U_{i}\int_{-\infty}^{\infty} dE\; f(E)\; Im{\bf
G}_{12}(i,i;E)
\label{eq:local order parameter}
\end{equation}
and the local band filling $n_{i}$ is,
\begin{equation}
 n_{i}=
2\int_{-\infty}^{\infty}dE\; f(E)\; Im{\bf G}_{11}(i,i;E) 
\label{eq:local band filling}
\end{equation}
where $f(E)=\frac{1}{exp({\frac{\beta E}{k_{B}T}})+1}$ is the Fermi function. Eq.\ref{eq:Bogoliobovdegenneequation} can be 
written as,
\begin{equation}
{\bf G}(i,j;E)={\bf G}^{0}(i,j;E)+\sum_{l}{\bf G}^{0}(i,l;E)
{\bf V}_{l}{\bf G}(l,j;E)
\label{eq:expanding interms of random onsite potential}
\end{equation}
where the random potential matrix ${\bf V}_{l}$ is
\begin{equation}
{\bf V}_{l}=\left(\begin{array}{cc}
\varepsilon_{l}+\frac{1}{2}U_{l}n_{l} & \Delta_{l}
\\
\Delta^{*}_{l} &-\varepsilon_{l}-\frac{1}{2}U_{l}n_{l}
\end{array}\right),
\label{eq:random onsite potential d-wave}
\end{equation}
and the clean normal system Green function ${\bf G}^{0}(i,j;E)$ is given by,
\begin{equation}
{\bf G}^{0}(i,j;E)=\frac{1}{N}\sum_{\bf k}e^{{\bf k}.{\bf r}_{ij}}
\left(\begin{array}{cc}
E-\epsilon_{\bf k} & 0
\\
 0 & E+\epsilon_{\bf k}
\end{array}\right)^{-1} ,
\label{eq:clean d-wave}
\end{equation}
where  $\epsilon_{\bf k}=-\mu-\frac{1}{N}\sum_{ij}t^{0}_{ij}e^{{\bf k}.{\bf r}_{ij}}$ is the bulk band structure. In our numerical calculations hopping to the first and second nearest neighbors is considered and we neglected other hopping terms, so $\epsilon_{\bf k}=-\mu-2t(cos(k_{x}a)+cos(k_{y}a))+t^{'}cos(k_{x}a)cos(k_{y}a)$.  The interaction potential is chosen to be $U_{i}=2t$. 

The Dyson equation for the average Green function ,$\bar{\bf G}(i,j;E)$ , corresponding to Eq.\ref{eq:expanding interms of random onsite potential} is,
\begin{eqnarray}
\bar{\bf G}(i,j;E)&=&{\bf G}^{0}(i,j;E)\nonumber\\&+&\sum_{ll^{'}}{\bf G}^{0}(i,l;E)
{\bf \Sigma}(l,l^{'},E)\bar{\bf G}(l^{'},j;E),
\label{eq:Dyson equation}
\end{eqnarray}
where the self energy, ${\bf \Sigma}(l,l^{'},E)$, is defined by,
\begin{equation}
\langle {\bf V}_{l}{\bf G}(l,j;E)\rangle=\sum_{l^{'}}{\bf \Sigma}(l,l^{'},E)\bar{\bf G}(l^{'},j;E).
\label{eq:self energy definition}
\end{equation}
The Fourier transformation of Eq.\ref{eq:Dyson equation} leads to the following relation for the average Green function $\bar{\bf G}(i,j;E)$,
\begin{eqnarray}
&\bar{\bf G}(i,j;E)=\frac{1}{N}\sum_{\bf k}e^{\imath{\bf k}.{\bf r}_{ij}}&
\nonumber\\\times&\left(\begin{array}{cc}
E-\epsilon_{\bf k} -{\Sigma}_{11}({\bf k},E)& -{\Sigma}_{12}({\bf k},E)
\\
 -{\Sigma}_{21}({\bf k},E) & E+\epsilon_{\bf k}-{\Sigma}_{22}({\bf k},E)
\end{array}\right)^{-1}.& \nonumber\\
\label{eq:avegrage Green function}
\end{eqnarray}
Eq.\ref{eq:local order parameter}-\ref{eq:avegrage Green function} should be solved to determine the averaged Green function, $\bar{\bf G}(i,j;E)$. In general there is no analytical solution for such systems, so it should be solved approximately. Many approximations such as self-consistent Born approximation or Abrikosov-Gorkov theory where is valid for weak scattering ($\delta\;\langle\langle\;band\;width$) and all impurity concentrations\cite{Abrikosov:59}, self-consistent T-matrix approximation where is limited to the case of low impurity concentrations\cite{Moradian2:00} and the coherent potential approximation (CPA) where is valid for all impurity concentrations and impurity strengths\cite{Moradian1:01}, $\delta$, and also recover both self-consistent Boron approximation (SCBA) and self-consistent T-matrix approximation (SCTMA) \cite{martin:99, Moradian2:00}, have been applied. To compare Abrikosov-Gorkov theory (weak scattering limit) and strong scattering case ($\delta\;\rangle\rangle\;band\;width$) we use CPA where in the weak impurity strengths ($\delta\;\langle\langle\;band\;width$) is equal to the self-consistent Boron approximation of Abrikosov-Gorkov theory.  


\section{Calculation of $T_{c}$ in the CPA formalism. }

In the CPA formalism inter-site correlations are neglected and each lattice site is replaced with a site of an effective medium except one which is called the {\em impurity site} and is denoted by $i$. So the self energy is local and takes the same value for all sites, ${\bf \Sigma}(l,m,E)={\bf \Sigma}(E)\delta_{lm}$, hence Eqs.\ref{eq:self energy definition} and \ref{eq:avegrage Green function} are reduce to\cite{Moradian2:00},
\begin{equation}
\langle {\bf V}_{i}{\bf G}^{imp}(i,i;E)\rangle={\bf \Sigma}(E)\bar{\bf G}(i,i;E)
\label{eq:cpa self energy definition}
\end{equation}
and
\begin{eqnarray}
\bar{\bf G}(l,m;E)=\frac{1}{N}\sum_{\bf k}e^{{\bf k}.{\bf r}_{lm}}
\nonumber\\\left(\begin{array}{cc}
E-\epsilon_{\bf k} -{\Sigma}_{11}(E)& -{\Sigma}_{12}(E)
\\
 -{\Sigma}_{21}(E) & E+\epsilon_{\bf k}-{\Sigma}_{22}(E)
\end{array}\right)^{-1} \nonumber\\
\label{eq:cpa avegrage Green function}
\end{eqnarray}
respectively, where the {\em impurity} Green function, ${\bf G}^{imp}(i,i,E)$, is related to the averaged Green function $\bar{\bf G}(i,i,E)$ as\cite{Moradian2:00},
\begin{eqnarray}
{\bf G}^{imp}(i,i;E)&=&\bar{\bf G}(i,i;E)\nonumber\\
&+&\bar{\bf G}(i,i;E)({\bf V}_{i}-{\bf\Sigma}(E)){\bf G}^{imp}(i,i;E).\nonumber\\
\label{eq:impurity Green function}
\end{eqnarray}
The new average Green function $\bar{\bf G}(i,i;E)$ is given by taking an average over all possible impurity configurations,
\begin{eqnarray}
\bar{\bf G}(i,i;E)=\langle{\bf G}^{imp}(i,i;E)\rangle
\label{eq:avimpurity Green function}
\end{eqnarray}
Eqs.\ref{eq:cpa self energy definition}- \ref{eq:avimpurity Green function} should be solved self consistently to provide $\bar{\bf G}(i,i;E)$. We continue our discussions by derivation of a set of equations to obtain the critical temperature in the CPA formalism. 

At $T_{c}$ both the local and average order parameter $\Delta_{i}$ and ${\bar\Delta}$  go to zero, and the spatial deviation of the local order parameter, $\delta\Delta_{i}$, with respect to the averaged order parameter, ${\bar\Delta}$, is negligible, so
\begin{equation}
\Delta_{i}\approx {\bar \Delta}.
\label{eq:cpa equality of order parameters}
\end{equation}
By inserting Eq.\ref{eq:cpa equality of order parameters} in Eq.\ref{eq:cpa self energy definition} we found that,
\begin{eqnarray}
\langle (\varepsilon_{i}-\frac{1}{2}U_{i} n_{i})G_{11}(i,i;E)\rangle&+& {\bar\Delta}{\bar G}_{21}(i,i;E)=\nonumber\\{ \Sigma}_{11}(E)\bar{ G}(i,i;E)&+&{ \Sigma}_{12}(E)\bar{ G}_{21}(i,i;E).
\label{eq:Tc cpa}
\end{eqnarray}
Since at $T_{c}$ the system is normal, the first terms in the left hand side and right hand side of Eq.\ref{eq:Tc cpa} are equal and cancel each other. Therefore we find,
\begin{equation}
{ \Sigma}_{12}(E)={\bar \Delta}.
\label{eq:cpa self energy and order parameter}
\end{equation}
By inserting Eq.\ref{eq:cpa self energy and order parameter} into the gap equation, Eq.\ref{eq:local order parameter}, for the average order parameter,$\bar\Delta$, at $T_{c}$ we obtain,
\begin{eqnarray}
1&=&U_{i}\int_{-\infty}^{\infty} dE\; f(E)\;\times \nonumber\\&Im&(\int_{-\infty}^{\infty}d\epsilon \frac{N_{0}(\epsilon)}{(E-\epsilon-\Sigma_{11}(E)(E+\epsilon-\Sigma_{22}(E))})
\nonumber\\
\label{eq:Tc equation}
\end{eqnarray}
 where $N_{0}(\epsilon)$ is the clean system density of states defined by,
\begin{equation}
N_{0}(\epsilon)=\frac{1}{N}\sum_{\bf k}\delta(\epsilon-\epsilon_{\bf k}).
\label{eq:clean density of states}
\end{equation}
 Eqs.\ref{eq:cpa equality of order parameters} and \ref{eq:cpa self energy and order parameter},
 Eqs.\ref{eq:cpa self energy definition}-\ref{eq:avimpurity Green function} should be solved self consistently to calculate $\Sigma_{11}(E)$ and $\Sigma_{22}(E)$. Then these self energies should be inserted in to Eq.\ref{eq:Tc equation} to obtain the critical temperature, $T_{c}$. 

\section{Results and Discussion}

\begin{figure}
\centerline{\epsfig{file=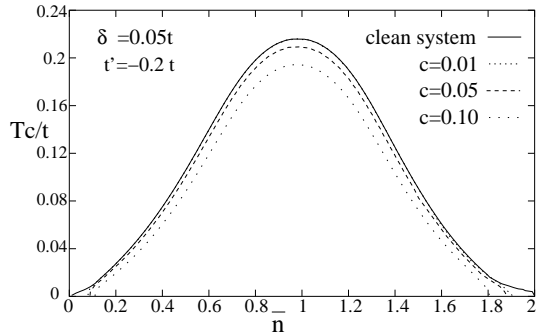  ,width=7.0cm,angle=0}}
\caption{ $T_{c}$ in terms of average band filling in the weak scattering regime (SCBA or Abrikosov-Gorkov theory) for different impurity concentrations, $c$, where $\delta=0.05 t\; \langle\langle \; bandwidth$. The deviation of $T_{c}$ from the clean system critical temperature, $T_{c0}$, is small. Hence AT is satisfied in the weak scattering and low impurity concentration regime. 
 \label{figure:Tc-delta-c0.05} }
\end{figure}

We considered several cases, first in the weak scattering regime, $\delta=0.05 t$ where CPA is equal to SCBA ( Abrikosov-Gorkov theory), $T_{c}$ is calculated in terms of average band filling, $\bar n$, for different impurity concentrations. Fig.\ref{figure:Tc-delta-c0.05} shows $T_{c}$ in terms of $\bar n$ at a fixed impurity strength, $\delta=0.05t$, for different impurity concentrations. As one can see, $T_{c}$ is changed only very little with respect to the clean system, and hence Anderson's theorem is satisfied in weak scattering, as expected.

\begin{figure}
\centerline{\epsfig{file= 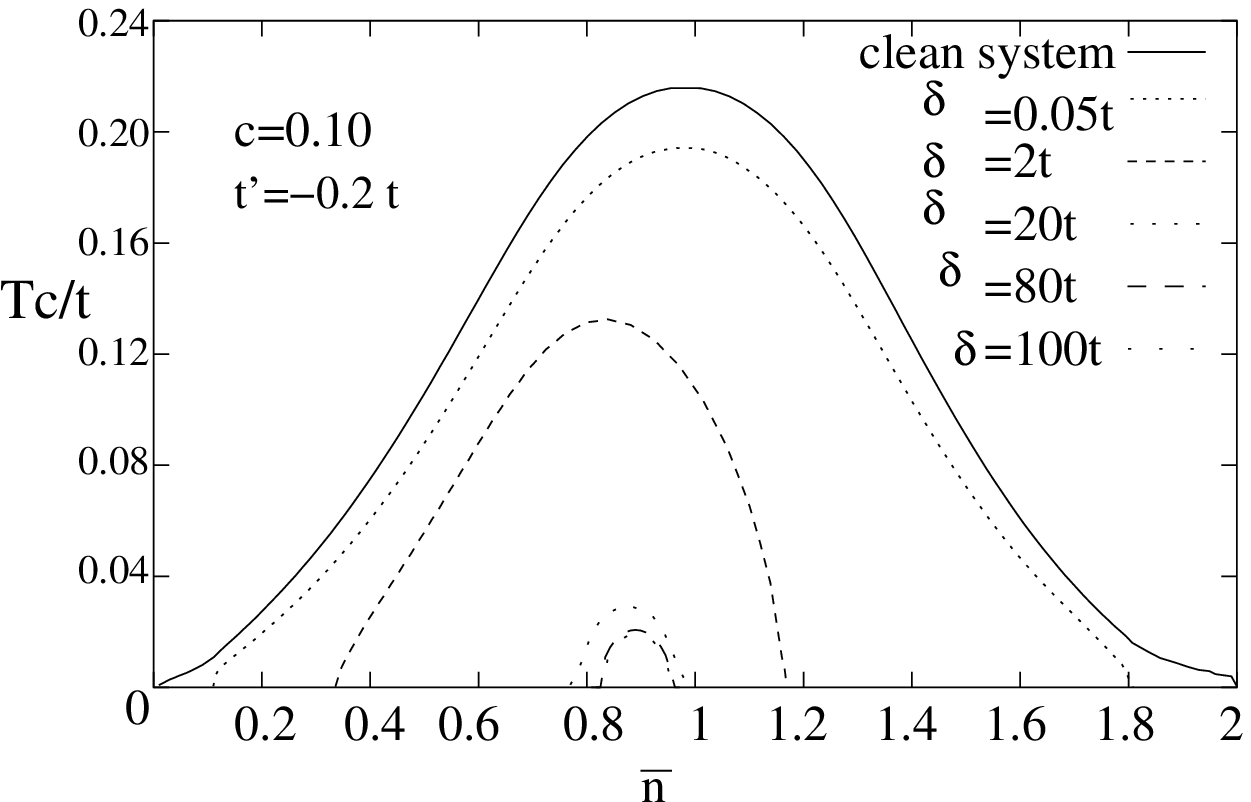,width=7.0cm,angle=0}}
\centerline{\epsfig{file=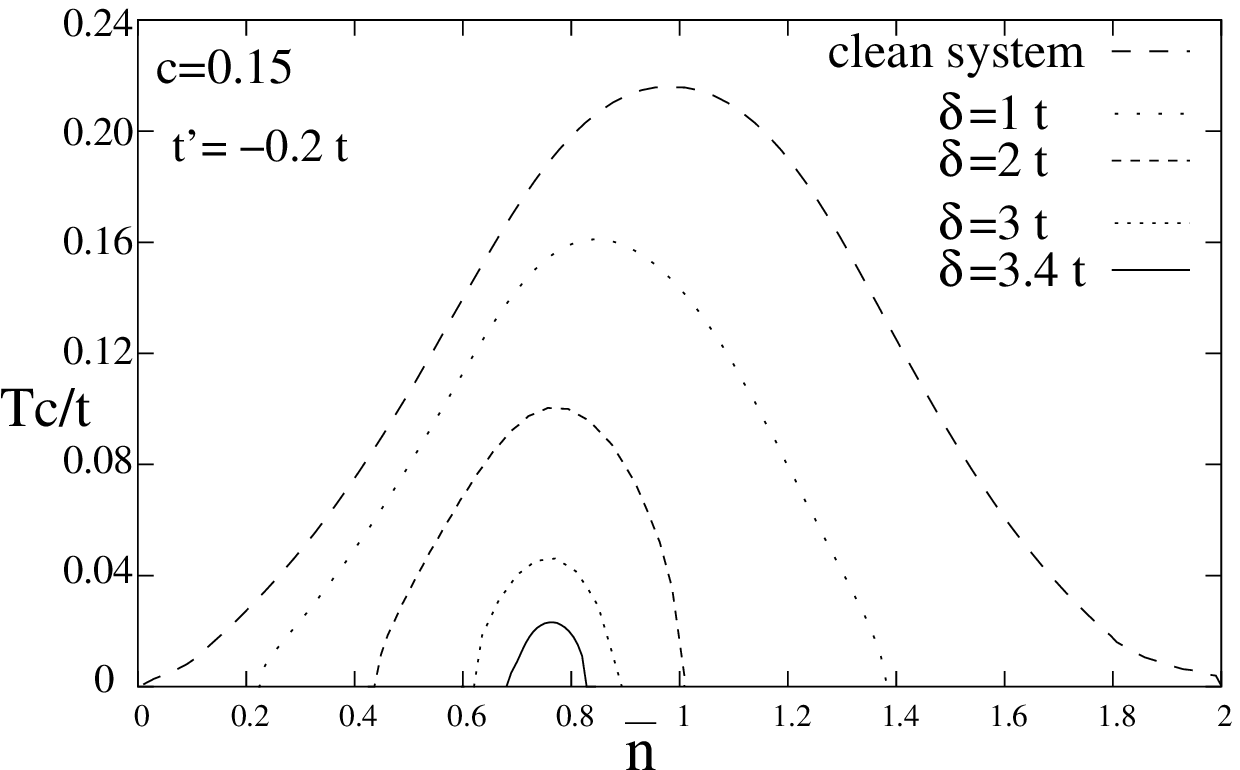 ,width=7.0cm,angle=0}}
\centerline{\epsfig{file= 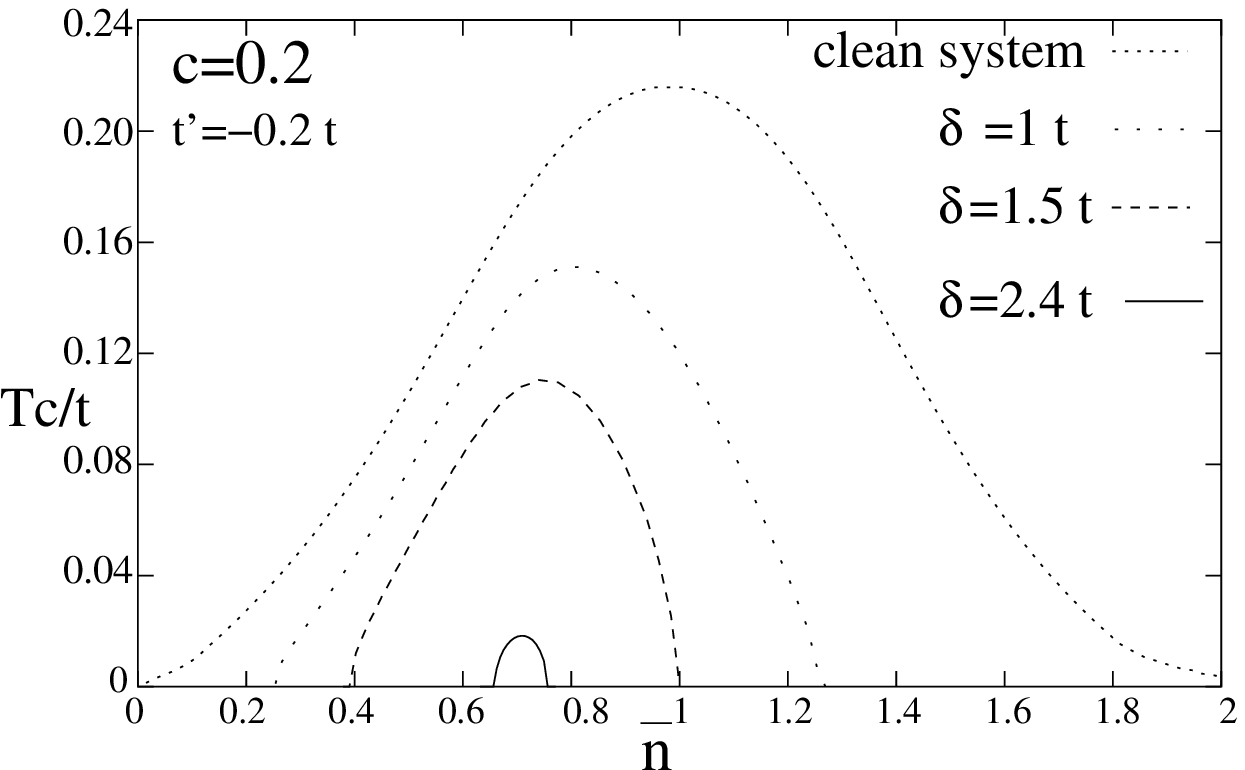,width=7.0cm,angle=0}}
\caption{$T_{c}$ for $c=0.1$, $c=0.15$ and $c=0.2$ in terms of band filling for different $\delta$ varying from weak (SCBA or Abrikosov-Gorkov theory) to strong scattering. In all these strong scattering cases $T_{c}$  is sensitive with respect to $c$ and $\delta$, so AT is not valid. For $c=0.1$ superconductivity is preserved even in $\delta=100t$ but for $c=0.15$ and $c=0.2$ superconductivity is suppressed at a critical $\delta$, so a superconductor-insulator quantum phase transition is take place.
 \label{figure:Tc-delta-c0.1-c.15-c.2} }
\end{figure}

 Second, $T_{c}$ was calculated in terms of $\bar n$ at fixed impurity concentrations, $c=0.1$, $c=0.15$ and $c=0.2$, for different value of $\delta$ varying from weak (SCBA or Abrikosov-Gorkov theory) to strong scattering. Fig.\ref{figure:Tc-delta-c0.1-c.15-c.2} illustrates our result in this case. For all these impurity concentrations in the strong scattering regime $T_{c}$ decreased with respect to the clean system significantly and for some band fillings there is no superconductivity. Also we found that for high impurity concentrations, $c=0.15$ and $c=0.2$ a superconductor-insulator quantum phase transition is take placed as previously predicted by us\cite{Moradian:01}   

\begin{figure}
\centerline{\epsfig{file=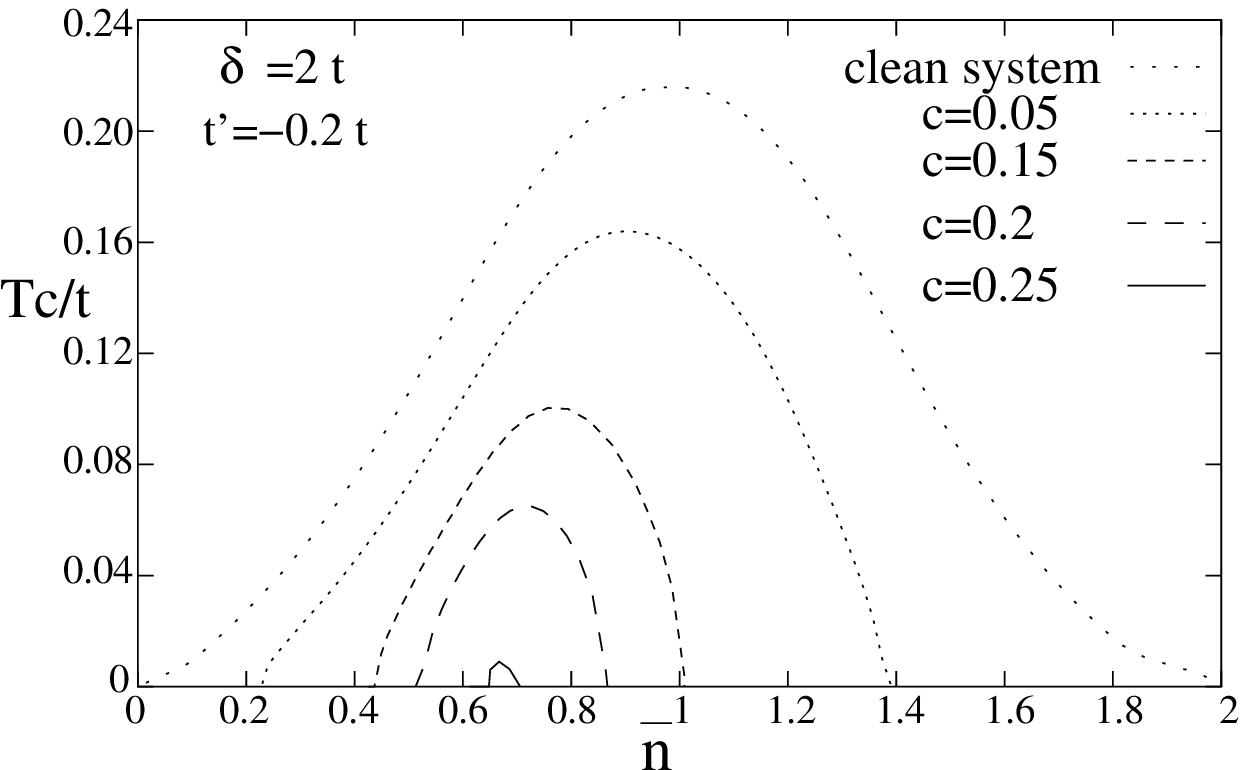 ,width=7.0cm,angle=0}}
\centerline{\epsfig{file=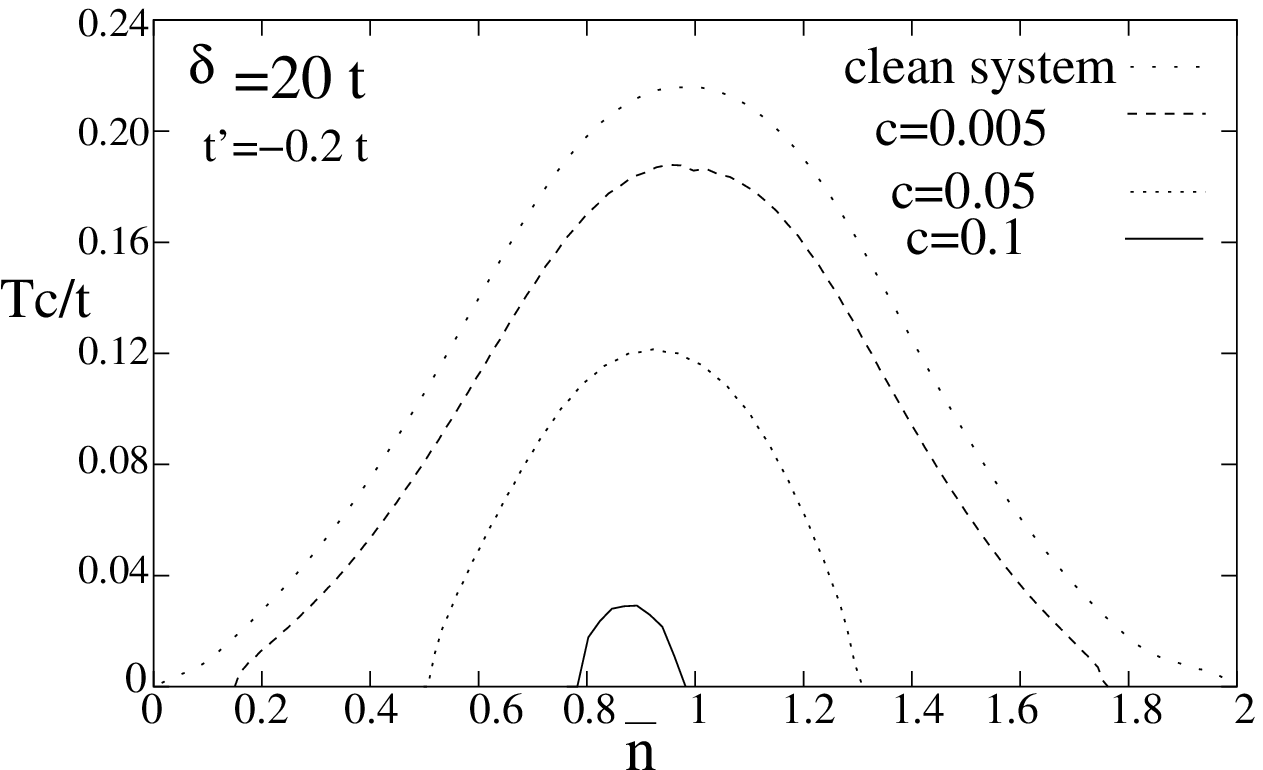 ,width=7.0cm,angle=0}}
\caption{ $T_{c}$ in terms of average band filling at fixed on site energies $\delta=2 t$ and  $\delta=20 t$ for different impurity concentrations. By increasing impurity concentration not only is $T_{c}$ reduced, but also
at a critical impurity concentration superconductivity is completely suppressed, hence a superconductor-insulator quantum phase transition has occurred. 
\label{figure:Tc-avfill-delta2t-delta20t-cpa} }
\end{figure}

Thirdly, $T_{c}$ was calculated for fixed values of the scattering strengths, $\delta=2t$ and $\delta=20t$, and different values of impurity concentration. We found by increasing impurity concentration $T_{c}$ is reduced and at a critical impurity concentration superconductivity completely disappears. Fig.\ref{figure:Tc-avfill-delta2t-delta20t-cpa} shows the results in this case.

\begin{figure}
\centerline{\epsfig{file=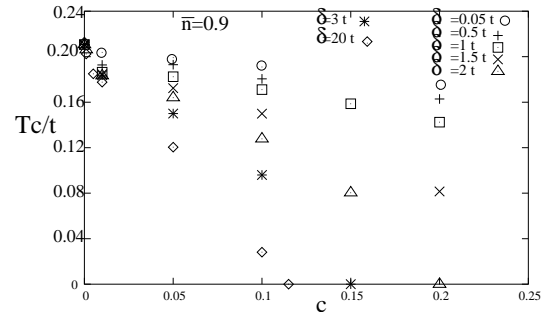, width=7.0cm,angle=0}}
\caption{ $T_{c}$ in terms of impurity concentration, $c$, for different values of scattering strength, $\delta$, where the average band filling is fixed at $\bar n=0.9$. For small $\delta$ (SCBA or Abrikosov-Gorkov theory regime) $T_{c}$ is not influenced
much and AT is valid, but in the strong scattering regime $T_{c}$ is reduced linearly by increasing impurity concentration until $T_c$ eventually becomes zero.
 \label{figure:Tc-c-nav0.9-cpa} }
\end{figure}

Finally, at a fixed band filling, $\bar n=0.9$, the critical temperature $T_{c}$ is plotted as a function of impurity concentration for different value of $\delta$.  Fig.\ref{figure:Tc-c-nav0.9-cpa} shows that for weak scattering $\delta\;\langle\langle\;band\;width$ (SCBA or Abrikosov-Gorkov theory), the critical temperature $T_{c}$ is not much influenced by increasing impurity concentration while in the strong scattering regime $T_{c}$ is decreases linearly with impurity concentration and superconductivity is completely destroyed.

 In conclusion, we have investigated response of the critical temperature $T_c$ of an s-wave superconductor to non-magnetic impurity doping. Three cases are considered. First, in the weak scattering regime $\delta\;\langle\langle\;band\;width$ (SCBA or Abrikosov-Gorkov theory regime),  $T_{c}$ is calculated for different impurity concentrations.
Here  we found that the $T_{c}$ is not affected by impurity doping so the Anderson theorem is satisfied in the case of low impurity concentrations and weak scattering. Second, at fixed impurity concentrations $c=0.1$, $c=0.15$ and $c=0.2$, $T_{c}$ is calculated in terms of band filling for different value of $\delta$. We found that by increasing $\delta$ the critical temperature decreased and at a critical $\delta$ a superconductor-insulator quantum phase transition takes place. Third, at fixed $\delta=2t$ and $\delta=20t$ the critical temperature is calculated in terms of band filling. Our results in this case show that by increasing impurity concentration $T_{c}$ is again reduced, and at a critical concentration a superconductor-insulator quantum phase transition again takes place. Finally at a fixed band filling $\bar n=0.9$, $T_{c}$ is plotted in terms of impurity concentration. In this case we found in the weak scattering regime (SCBA or Abrikosov-Gorkov theory regime) $T_{c}$ is not much influenced by increasing impurity concentration but at strong scattering regime $T_{c}$ is linearly decreased by increasing impurity concentration.  Therefore, we conclude that Anderson's theorem\cite{Anderson:59} 
only holds in the weak scattering (Born approximation) regime, originally considered by Abrikosov and Gorkov\cite{Abrikosov:59}. In stronger scattering even non-magnetic impurities can suppress s-wave superconductivity
and lead to greatly reduced $T_c$ or even a transition to a non-superconducting insulator state.

\acknowledgements We would like to thanks James F. Annett and B. L. Gyorffy for helpful discussions.

\end{document}